\documentclass[notitlepage,12pt]{iopart}

\usepackage{epsf}
\begin{document}

\title[]{Dissipative dynamics of coupled quantum dots under quantum measurement}

\author{Shi-Hua Ouyang$^1$, Chi-Hang Lam$^2$, J Q You$^1$}

\address{$^1$Department of Physics
and Surface Physics Laboratory (National Key Laboratory), Fudan
University, Shanghai 200433, China}
\address{$^2$Department of Applied Physics, Hong Kong Polytechnic
University, Hung Hom, Hong Kong, China}
\date{\today}
\begin{abstract}
  We have studied the dissipative dynamics of a solid-state qubit with
  an extra electron confined to either one of two coupled quantum
  dots. Previous theoretical work based on Bloch-type rate equations gave
  an unphysical uniform occupation probability of the electron in
  the quantum dots even for non-identical dots. We show that this is
  due to neglecting higher order interactions in the analysis. By
  including higher order terms, we obtain expected asymmetric
  occupation probabilities for non-identical dots. Our work
  demonstrates that the high order interaction terms can lead to
  important qualitative impacts on the operation of the qubit.
\end{abstract}
%
\section{Introduction}
Quantum measurement has attracted
much attention because of its fundamental importance as well as its
relevance to quantum computing. However, there are still vastly
different points of view on quantum measurement. On one hand, the
Zeno effect predicts that continuous measurement can freeze the
state of a quantum system~\cite{ Namiki97}. On the other hand,
interaction with the outside environment such as a detector is
expected to lead to the collapse of the state of the measured
system. Therefore, reconciling these views by detailing the real
mechanisms for selected model systems is very important.

Many experiments on two-level quantum systems (see, e.g.,
Refs.~\cite{Yacoby94,Shuster97,Buks96}) have been performed to study
this problem. Together with experimental investigations, some
theoretical studies were also devoted to this important topic.
Gurvitz {\it et al}~\cite{Gurvitz97,Gurvitz03} have theoretically
studied a mesoscopic qubit-plus-detector system. The qubit consists
of two coupled-quantum dots (CQD's) while the detector is a quantum
point contact (QPC) interacting locally with one of the dots. The
authors derived a set of Bloch-type rate equations to describe the
measurement effects. Goan {\it et al}~\cite{Goan01} considered the
same setup and developed a Lindblad quantum master equation using
the quantum trajectory approach for dissipative processes. However,
while reasonable result was obtained for a pair of identical quantum
dots in both Refs.~\cite{Gurvitz97} and \cite{Goan01}, these works
predicted unreasonable equal occupation probabilities in the dots
when their energy levels are different. Subsequently, Li {\it et
al}~\cite{Xinqi04} developed a unified quantum master equation with
the Markovain approximation and obtained distinct occupation
probabilities for non-identical quantum dots as expected.

In this work, we further study this qubit-plus-detector system
following the original approach of Gurvitz {\it et
al}~\cite{Gurvitz97}. We have found that the unphysical result of
identical occupation probabilities in non-identical dots mentioned
above is only due to neglecting higher-order terms in their
derivation. We will show that higher-order terms are of fundamental
importance. Retaining additional terms in our calculations leads to
physically valid results consistent with those of Li {\it et
al}~\cite{Xinqi04}. Our analysis also illustrates the detailed
mechanism of quantum measurement.

\section{Bloch-type rate equations}

\noindent The system we study is shown schematically in Fig. 1. It
can be modeled by the Hamiltonian
$H=H_{\rm{CQD}}+H_{\rm{QPC}}+H_{\rm{int}}$, with
%
\begin{eqnarray}
H_{\rm{CQD}}\!&\!=\!&\!
E_1a_1^{\dag}a_1+E_2a_2^{\dag}a_2+\Omega_0(a_1^{\dag}a_2+a_2^{\dag}a_1),
\nonumber\\
H_{\rm{QPC}}\!&\!=\!&\!
\sum_{l}E_la_l^{\dag}a_l+\sum_{r}E_ra_r^{\dag}a_r
+\sum_{l,r}\Omega_{lr}(a_l^{\dag}a_r+\rm{H.c.}),
\nonumber\\
H_{\rm{int}}\!&\!=\!&\!
\sum_{l,r}\delta\Omega_{lr}a_1^{\dag}a_1(a_l^{\dag}a_r+a_r^{\dag}a_l).\label{eq1}
\end{eqnarray}
Here, $H_{\rm{CQD}}$ is the Hamiltonian of the CQD's. The Hubbard
terms are neglected because we consider only one extra electron in
the CQD's. The parameter $\Omega_0$ denotes the hopping amplitude of
the extra electron between the two single-dot states (see Fig.1).
$H_{\rm{QPC}}$ is the Hamiltonian of the QPC, and $H_{\rm{int}}$ is
the interaction Hamiltonian between the CQD's and the QPC.
Electrostatic effects are included in $H_{\rm{int}}$: an extra
electron in the left dot will lead to an effective variation
$\delta\Omega_{lr}$ in the coupling $\Omega_{lr}$ between the states
$E_l$ and $E_r$ in the two reservoirs of the QPC. The resulting
coupling then becomes $\Omega'_{lr}= \Omega_{lr} +
\delta\Omega_{lr}$.

\begin{figure} \epsfxsize 6.0cm
\centerline{\epsffile{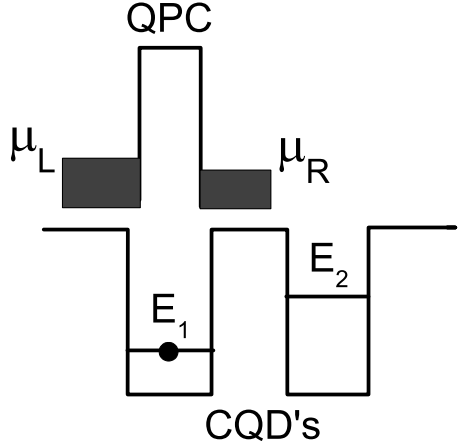}} \caption{Two coupled quantum dots
(CQD's) connected with a quantum
point contact (QPC). Initially, all states up to the Fermi energy level
($\mu_L$ or $\mu_R$) in each reservoir are filled by electrons.
$E_1$ and $E_2$ denote the energy levels of an extra electron in
the left and right quantum dots, respectively. The location of this
extra electron will affect the current through the QPC. The quantum
state of the CQD's can hence be read out by a current measurement.}
\end{figure}

In the occupation representation, the many-body wave function of the
entire system (the CQD's plus the QPC) can be written as
%
\begin{eqnarray}
|\psi(t)\rangle = [b_1(t)a_1^{\dag}+b_2(t)a_2^{\dag}
\!+\sum_{l,r}b_{1lr}(t)a_1^{\dag}a_r^{\dag}a_l+\sum_{l,r}b_{1rl}(t)a_1^{\dag}a_l^{\dag}a_r
\nonumber\\
+\sum_{l,r}b_{2lr}(t)a_2^{\dag}a_r^{\dag}a_l+\sum_{l,r}b_{2rl}(t)a_2^{\dag}a_l^{\dag}a_r
+\sum_{l\rm{<}l',r<r'}b_{1ll'rr'}(t)
a_1^{\dag}a_r^{\dag}a_{r'}^{\dag}a_la_{l'} \nonumber\\
+\sum_{l\rm{<}l',r<r'}b_{1rr'll'}(t)
a_1^{\dag}a_l^{\dag}a_{l'}^{\dag}a_ra_{r'}
+\sum_{l<l',r<r'}b_{2ll'rr'}(t)
a_2^{\dag}a_r^{\dag}a_{r'}^{\dag}a_la_{l'}\nonumber\\
+\sum_{l<l',r<r'}b_{2rr'll'}(t)
a_2^{\dag}a_l^{\dag}a_{l'}^{\dag}a_ra_{r'}+\cdots
]|0\rangle\,,\label{eq2}
\end{eqnarray}
%
where $b_j(t)$, $j=1, 2, 1lr, 2lr$, ${\cdots}$, are the
time-dependent probability amplitudes of finding the system in the
corresponding states.
The vacuum state $|{0}\rangle$ corresponds to the state of the
entire system with no extra electron in the CQD's and before the
transfer of any electron between the reservoirs in the QPC.
Then, $b_{1lr}(t)$ for instance denotes the probability amplitude
that the extra electron is in the left dot and an electron has
passed through the barrier in the QPC.  The initial conditions are
$b_1(0)=1$ and $b_j(0)=0$ for $j=2, 1lr, 2lr$, ${\cdots}$.
Substituting the many-body wave function [Eq. (\ref{eq2})] and the
Hamiltonian [Eq. (\ref{eq1})] into the time-dependent
Schr\"{o}dinger equation
\begin{math}i|\dot{\psi}(t)\rangle=H|\psi(t)\rangle\end{math} gives
an infinite set of linear equations for the probability amplitudes
$b_j(t)$. Performing a Laplace transform to each equation, the
transformed amplitude
%
\begin{math}
b_j(E)=\int_{0}^{+\infty}e^{iEt}b_j(t)dt
\end{math}
~follows:
\numparts
\begin{eqnarray}
\fl (E-E_1)b_1(E)-\Omega_0b_2(E)-\sum_{l,r}\Omega'_{lr}b_{1lr}(E)-\sum_{l',r'}\Omega'_{l'r'}b_{1r'l'}(E)=i,\label{eq3a}\\
\fl (E-E_2)b_2(E)-\Omega_0b_1(E)-\sum_{l,r}\Omega_{lr}b_{2lr}(E)-\sum_{l',r'}\Omega_{l'r'}b_{2r'l'}(E)=0,\label{eq3b}\\
\fl (E+E_l-E_1-E_r)b_{1lr}(E)-\Omega'_{lr}b_1(E)-\Omega_0b_{2lr}(E)
-\sum_{l'r'}\Omega'_{l'r'}b_{1ll'rr'}(E)=0,\label{eq3c}\\
\fl (E+E_l-E_2-E_r)b_{2lr}(E)-\Omega_{lr}b_2(E)-\Omega_0b_{1lr}(E)
-\sum_{l'r'}\Omega_{l'r'}b_{2ll'rr'}(E)=0,\label{eq3d}\\
\fl (E+E_r-E_1-E_l)b_{1rl}(E)-\Omega'_{lr}b_1(E)-\Omega_0b_{2rl}(E)
-\sum_{l'r'}\Omega'_{l'r'}b_{1rr'll'}(E)=0,\label{eq3e}\\
\fl (E+E_r-E_2-E_l)b_{2rl}(E)-\Omega_{lr}b_2(E)-\Omega_0b_{1rl}(E)
-\sum_{l'r'}\Omega_{l'r'}b_{2rr'll'}(E)=0,\label{eq3f}\\
\lefteqn{{\cdots}{\cdots}{\cdots}.}\nonumber\hspace{5cm}
\end{eqnarray}
\endnumparts
To solve Eqs.~({3}), Gurvitz {\it et al.}~\cite{Gurvitz97,Gurvitz03}
introduced an approximation by keeping terms up to the order of
$\Omega^2$ to simplify the corresponding sum terms [see Eqs.~(A5)
in Ref.~5]. After tracing out the degrees of freedom of QPC, a
coupled set of Bloch-type rate equations are derived. As shown in
Fig.~3 of Ref.~5(b), however, the resulting occupation probabilities
for states $|E_1\rangle$ and $|E_2\rangle$ are 0.5 when
$t\rightarrow\infty$ for both symmetric and asymmetric CQD's. This
is correct for the symmetric case with identical quantum dots, but
is unphysical in the latter case.
%
%
Here, we show that this unphysical feature can be rectified by
taking terms of the order of $\Omega^2\Omega_0/V_d$ or even higher into
account. Also, following Ref.~5, we consider the high voltage regime
with $eV_d\gg\Omega^2\rho$, where $V_d=\mu_L-\mu_R$ is the applied
gate voltage, and $\rho$ is the density of states in the reservoirs.
In this high voltage regime, when the temperature is low enough (here it is 
chosen zero for simplicity, as in Ref.~5), the electron has an extremely 
low probability to pass through the QPC from the right reservoir (with a 
lower $\mu_R$) to the left one (with a higher $\mu_L$). 
Thus, the terms describing the back processes, e.g.,
$\sum_{l\rm{<}l',r<r'}b_{1rr'll'}(t)$, can be neglected.

Taking the sum term
$\sum_{l,r}\Omega'_{lr}b_{1lr}(E)$ in Eq.~(\ref{eq3a}) for example,
we obtain that
\begin{eqnarray}
\sum_{l,r}\Omega'_{lr}b_{1lr}(E)\approx-\frac{iD'}{2V_d}(V_d+E-E_1)b_1(E)+\frac{i\Lambda}{2V_d}b_2(E),\label{eq4}
\end{eqnarray}
where $D'=2\pi\rho_l\rho_r{\Omega'}^2V_d$,
$\Lambda=2\pi\rho_l\rho_r\Omega_0\Omega\Omega'V_d$, and the
corresponding exact calculation are shown in the Appendix. With the
same approximate treatment, the resulting equations for $b_j(E)$ are
\numparts
\begin{eqnarray}
\fl [E-E_1+\frac{iD'}{2V_d}(V_d+E-E_1)]b_1(E)
-(\Omega_0+\frac{i\Lambda}{2V_d})b_2(E)\!=\!i, \label{eq5a}
\\
\fl [E-E_2+\frac{iD}{2V_d}(V_d+E-E_2)]b_2(E)
-(\Omega_0+\frac{i\Lambda}{2V_d})b_1(E)=0,\label{eq5b}
\\
\fl [E+E_{lr1}+\frac{iD'}{2V_d}(V_d+E+E_{lr1})]b_{1lr}(E)
-\Omega'b_1(E)-(\Omega_0+\frac{i\Lambda}{2V_d})b_{2lr}(E)=0,
\label{eq5c} \\
\fl [E+E_{lr2}+\frac{iD}{2V_d}(V_d+E+E_{lr2})]b_{2lr}(E)
-{\Omega}b_2(E)-\left(\Omega_0+\frac{i\Lambda}{2V_d}\right)b_{1lr}(E)=0,
\label{eq5d}
\\
%
%
%
~~~~~~~~~~~~~~~~~~~~\lefteqn{{\cdots}{\cdots}{\cdots},}\hspace{5cm}\nonumber
\end{eqnarray}
\endnumparts
%
where $D=2\pi\rho_l\rho_r\Omega{^2}V_d$ and $E_{lrm}=E_l-E_m-E_r$ ($m=1,2$).
Now, we introduce the notation
\begin{eqnarray*}
\sigma_{ij}^{(n)}=\sum\limits_{l{\cdots}r{\cdots}}
b_{il{\cdots}r{\cdots}}(t) b^*_{jl{\cdots}r{\cdots}}(t).
\end{eqnarray*}
For instance, $\sigma_{11}^{(1)}=\sum\limits_{lr}
b_{1lr}(t)b^*_{1lr}(t)$ denotes the occupation probability for the
extra electron staying in the left quantum dot and an electron passing
through the QPC.

The equations for the amplitudes $b_j(E)$, $j=1, 2, 1lr, 2lr$,
${\cdots}$, can be converted to a new set of equations for
$\sigma_{ij}^{(n)}$ using the inverse Laplace transform:
\begin{eqnarray*}
\sigma_{ij}^{(n)}=\sum\limits_{l{\cdots}r{\cdots}}\int\frac{dEdE'}{4\pi^2}
b_{il{\cdots}r{\cdots}}(E)b_{jl{\cdots}r{\cdots}}^*(E')e^{i(E'-E)t}.
\end{eqnarray*}
%
We now multiply Eq.~(\ref{eq5c}) by $b_{1lr}^{*}(E')$. The resulting
equation is then subtracted by its complex conjugate after
exchanging
$E$ with 
${E'}$. We obtain
\begin{eqnarray}
\fl
\int\int\frac{dEdE'}{4\pi^2}\sum_{lr}\{(E'-E-iD')b_{1lr}(E)b_{1lr}^{*}(E')
\nonumber\\
\fl
-\frac{iD'}{2V_d}[(E'+E_{lr1})+(E+E_{lr1})]b_{1lr}(E)b^*_{1lr}(E')
\nonumber\\
\fl
+\frac{i\Lambda}{2V_d}[b_{2lr}(E)b_{1lr}^{*}(E')+b_{1lr}(E)b_{2lr}^{*}(E')]
-\Omega_0[b_{1lr}(E)b_{2lr}^{*}(E')-b_{2lr}(E)b_{1lr}^{*}(E')]
\nonumber\\
\fl
-\Omega'[b_{1}^{*}(E')b_{1lr}(E)-b_{1}(E)b_{1lr}^{*}(E')]\}e^{i(E'-E)t}=0.\label{eq6}
\end{eqnarray}
Using Eqs.~4, (\ref{eq5a}) and (\ref{eq5c}), we can obtain
from Eq.~(\ref{eq6}), by neglecting terms of order $O(\Omega^6/V_d^2)$ and
higher, that
\begin{eqnarray}
\fl
\dot{\sigma}_{11}^{(1)}\!=\!-D'\big[\sigma_{11}^{(1)}-\sigma_{11}^{(0)}\big]
+i\Omega_0\big[\sigma_{12}^{(1)}-\sigma_{21}^{(1)}\big]
+\frac{\Lambda}{2V_d}\left(1-\frac{\Omega'}{\Omega}\right)\big[\sigma_{12}^{(1)}
+\sigma_{21}^{(1)}-\sigma_{12}^{(0)}-\sigma_{21}^{(0)}\big].\;\;\;\;\label{eq7}
\end{eqnarray}
Similarly, Eq.~(\ref{eq5d}) gives
\begin{eqnarray}
\fl \dot{\sigma}_{22}^{(1)}
\!=\!-D\big[\sigma_{22}^{(1)}-\sigma_{22}^{(0)}\big]
+i\Omega_0\big[\sigma_{21}^{(1)}-\sigma_{12}^{(1)}\big]
+\frac{\Lambda}{2V_d}\left(1-\frac{\Omega}{\Omega'}\right)
\big[\sigma_{12}^{(1)}+\sigma_{21}^{(1)}-\sigma_{12}^{(0)}
-\sigma_{21}^{(0)}\big].\;\;\;\;\label{eq8}
\end{eqnarray}
%
To calculate the off-diagonal element $\sigma_{12}$, we subtract the
product of Eq.~(\ref{eq5c}) and $b_{2lr}^{*}(E')$ by the product of
$b_{1lr}(E)$ and the complex conjugate of Eq.~(\ref{eq5d}) after exchanging
$E\leftrightarrow{E'}$. We obtain
\begin{eqnarray}
\fl \dot{\sigma}_{12}^{(1)}\!=\!i\varepsilon\sigma_{12}^{(1)}
-\frac{1}{2}(D'+D)\sigma_{12}^{(1)}+\sqrt{DD'}\sigma_{12}^{(0)}
+i\Omega_0\big[\sigma_{11}^{(1)}-\sigma_{22}^{(1)}\big]
-\frac{\Lambda}{2V_d}\left[\frac{\Omega'}{\Omega}\sigma_{11}^{(0)}
+\frac{\Omega}{\Omega'}\sigma_{22}^{(0)}\right]
\nonumber\\
\!\!
+\frac{\Lambda}{2V_d}\big[\sigma_{11}^{(0)}+\sigma_{22}^{(0)}\big]
-\frac{\Lambda}{2V_d}\left[\frac{\Omega}{\Omega'}\sigma_{11}^{(1)}
+\frac{\Omega'}{\Omega}\sigma_{22}^{(1)}\right]
+\frac{\Lambda}{2V_d}\big[\sigma_{11}^{(1)}+\sigma_{22}^{(1)}\big],\label{eq9}
\end{eqnarray}
where $\varepsilon=E_2-E_1$ is the energy-level difference of the two
quantum dots. Similar procedures can be used for $\sigma_{ij}^{(n)}$
($n\geq2$) and the resulting equations are
\begin{eqnarray}
\fl \dot{\sigma}_{11}^{(n)}\!=\!-D'\big[\sigma_{11}^{(n)}
-\sigma_{11}^{(n-1)}\big]+i\Omega_0\big[\sigma_{12}^{(n)}-\sigma_{21}^{(n)}\big]
+\textstyle\frac{\Lambda}{2V_d}\left(1-\frac{\Omega'}{\Omega}\right)\big[\sigma_{12}^{(n)}
+\sigma_{21}^{(n)}-\sigma_{12}^{(n-1)}-\sigma_{21}^{(n-1)}\big],
\nonumber\\
\fl \dot{\sigma}_{22}^{(n)}\!=\!-D\big[\sigma_{22}^{(n)}
-\sigma_{22}^{(n-1)}\big]+i\Omega_0\big[\sigma_{21}^{(n)}-\sigma_{12}^{(n)}\big]
\textstyle+\frac{\Lambda}{2V_d}\left(1-\frac{\Omega}{\Omega'}\right)\big[\sigma_{12}^{(n)}
+\sigma_{21}^{(n)}-\sigma_{12}^{(n-1)}-\sigma_{21}^{(n-1)}\big],
\nonumber\\
\fl \dot{\sigma}_{12}^{(n)}\!=\!i\varepsilon\sigma_{12}^{(n)}
-\frac{1}{2}(D'+D)\sigma_{12}^{(n)}+\sqrt{DD'}{\sigma_{12}^{(n-1)}}
+i\Omega_0\big[\sigma_{11}^{(n)}-\sigma_{22}^{(n)}\big]
+\textstyle\frac{\Lambda}{2V_d}\big[\sigma_{11}^{(n)}+\sigma_{22}^{(n)}\big]
\nonumber\\
\fl
-\textstyle\frac{\Lambda}{2V_d}\left[\frac{\Omega'}{\Omega}\sigma_{11}^{(n-1)}+\frac{\Omega}
{\Omega'}\sigma_{22}^{(n-1)}\right]
+\frac{\Lambda}{2V_d}\big[\sigma_{11}^{(n-1)}+\sigma_{22}^{(n-1)}\big]
-\frac{\Lambda}{2V_d}\left[\frac{\Omega}{\Omega'}\sigma_{11}^{(n)}
+\frac{\Omega'}{\Omega}\sigma_{22}^{(n)}\right]\,.\label{eq10}
\end{eqnarray}
Summing over $n$, we obtain the following Bloch-type rate equations
for the density-matrix elements:
\begin{eqnarray}
\dot{\sigma}_{11}\!&\!=\!&\! i\Omega_0(\sigma_{12}-\sigma_{21}),\nonumber\\
\dot{\sigma}_{22}\!&\!=\!&\! i\Omega_0(\sigma_{21}-\sigma_{12}),\nonumber\\
\dot{\sigma}_{12}\!&\!=\!&\!
i\varepsilon\sigma_{12}-\frac{\Gamma_d}{2}\sigma_{12}
+i\Omega_0(\sigma_{11}-\sigma_{22})
%
\! -\frac{\chi}{2}(\sigma_{11}+\sigma_{22}),\label{eq11}
\end{eqnarray}
with
\begin{eqnarray}
\Gamma_d \!= \!\big(\sqrt{D'}-\sqrt{D}\big)^2,
\chi \!=\!\Big(\frac{\Lambda}{V_d}\Big)
\Big(\frac{\Omega}{\Omega'}+\frac{\Omega'}{\Omega}-2\Big).\label{eq12}
\end{eqnarray}
%
Here $\sigma_{ij} =\sum_{n}\sigma_{ij}^{(n)}$ is the reduced
density-matrix elements of the CQD system after tracing out the
variables of the QPC. The decoherence rate $\Gamma_d$ characterizes
the exponential damping of the off-diagonal density-matrix element.
These Bloch-type rate equations are the same as those from a theory
based on master equations with the Markovain approximation
[cf.,~Eq.~(17) in Ref.~8]. 
When ignoring the effects of the higher-order terms $\chi=0$, these
equations are reduced to the rate equations derived in Ref.~5(b), 
which gives the unphysical result of identical occupation probabilities 
for asymmetrical CQD's.
This implies that the higher-order terms play an essential role in 
obtaining more accurate and reasonable results and help us better understand
the effects of the quantum measurement on the considered system.

Here we take into account the terms of the order 
$\Omega^2\Omega_0/V_d$, while these terms were ignored in Ref.~5. 
To obtain more accurate results, one can consider further higher-order terms, 
which have the order of $\Omega^6/V_d^2$ and $\Omega^6\Omega_0/V_d^3$. 
In the high voltage regime, these higher-order terms do not affect the results 
significantly (see the Appendix).     

\begin{figure}
\epsfxsize 9.0cm \centerline{\epsffile{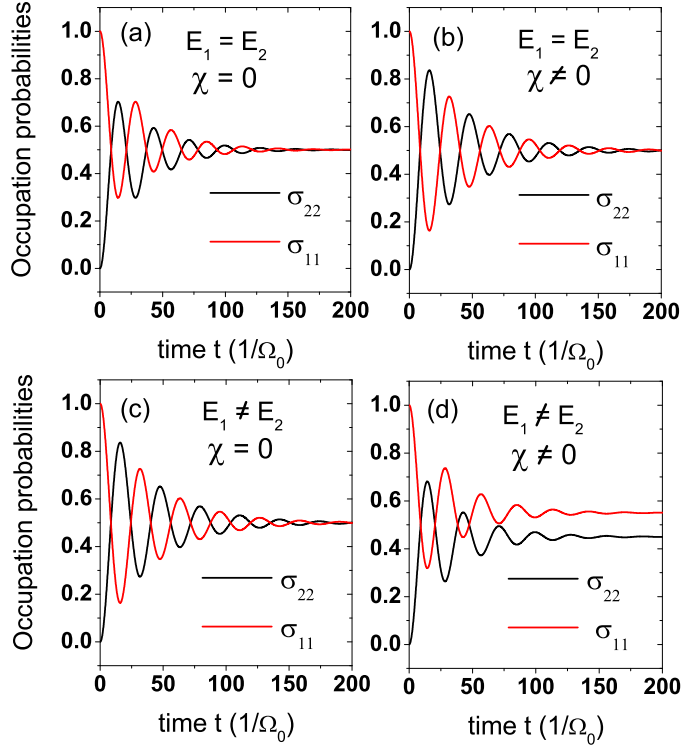}} \caption{(Color
online)~Occupation probability of the extra electron in each quantum
dot as a function of time for $\Gamma_d=\Omega_0$ and (a)
$\varepsilon=E_2-E_1=0$, $\chi=0$; (b) $\varepsilon=0$,
$\chi=0.1\Omega_0$; (c) $\varepsilon=\Omega_0$, $\chi=0$; and (d)
$\varepsilon=\Omega_0$, $\chi=0.1\Omega_0$. \label{fig2}}
\end{figure}

\section{Discussion and conclusion}
\noindent Figure 2 shows the time dependence of the occupation
probability of the extra electron in each quantum dot. When
$\chi=0$, the occupation probability in each dot always decays to
0.5 at $t\rightarrow\infty$ [see Figs.~(2a) and~(2c)]. These results
are identical to those in Ref.~5(b). However, after including higher
order terms, $\chi$ becomes nonzero and is not negligible. This
gives rise to an additional term
$\frac{1}{2}\chi(\sigma_{11}+\sigma_{22})$ in the last equation of
(11). From Fig.~(2b), one can see that the occupation probabilities
still decay to 0.5 as $t\rightarrow\infty$ for symmetrical CQD's,
but decay more slowly than in the previous case with vanishing
$\chi$. Moreover, for asymmetrical CQD's, the occupation probability
in each dot decays to a different value at $t\rightarrow\infty$ [see
Fig.~2(d)]. This is reasonable because the two quantum dots are
characterized by different parameters. Below we further explain these 
phenomena analytically. 

Let us look at the stationary solution of Eq.~(11) in the limit
$t\rightarrow\infty$ so that
$\dot\sigma_{ij}(t\rightarrow\infty)=0$. Using
$\sigma_{ij}(t\rightarrow\infty)=u_{ij}+iv_{ij}$, where $u_{ij}$ and
$v_{ij}$ are real numbers, Eq.~(11) can be rewritten as
\begin{eqnarray}
0\!&\!=\!&\!-2\Omega_0{v_{12}} \nonumber\\
0\!&\!=\!&\!2\Omega_0{v_{12}} \nonumber\\
0\!&\!=\!&\!i\varepsilon(u_{12}+iv_{12})
-\frac{\Gamma_d}{2}(u_{12}+iv_{12})
\!+i\Omega_0(\sigma_{11}-\sigma_{22})-\frac{\chi}{2}.
\end{eqnarray}
From these equations, one obtains
\begin{eqnarray}
\sigma_{11}-\sigma_{22}=\frac{\chi\varepsilon}
{\Omega_0\Gamma_d}=\frac{\varepsilon}{V_d}.
\end{eqnarray}
As expected, $\sigma_{11}=\sigma_{22}$ for $E_1=E_2$ (i.e.,
$\varepsilon=0$), while $\sigma_{11}\neq\sigma_{22}$ for
asymmetrical CQD's with $E_1\neq{E_2}$ (i.e., $\varepsilon\neq0$).

In conclusion, we have presented a quantitative description of the
dissipative dynamics of a CQD qubit connected to a detector in the
form of a QPC.  Bloch-type rate equations for the reduced density
matrix elements of the qubit are derived. In particular, the
long-time probability distribution of the state of the qubit is
found to depend on the energy levels of the single dot electron
states. This corrects an earlier calculation which has predicted a
distribution independent of the dot properties. The improvement
results from taking into account higher order interaction terms in
our analysis and it can be extended to other quantum systems. 
In the present work, the Bloch-type rate equations are derived 
for a CQD system working at zero temperature and applied in the strong 
bias voltage regime. However, a finite temperature will affect the asymptotic 
population of the qubit and a popoulation inversion can even be 
achieved in some cases~\cite{Goorden04}.

\section{Appendix:~Derivation of equations for probability amplitudes}
\appendix\setcounter{section}{1}

For the QPC, we assume that the hopping amplitude
$\Omega_{lr}(E_l,E_r)$ between the left and right reservoirs depends
only weakly on the energy levels $E_l$ and $E_r$. The energy
dependence is therefore neglected, i.e.
$\Omega_{lr}(E_l,E_r)=\Omega$, and
${\Omega'}_{lr}(E_l,E_r)=\Omega'$. As high-order terms
$\sum_{l'r'}{\Omega'}b_{1ll'rr'}(E)$ and $
\sum_{l'r'}{\Omega}b_{2ll'rr'}(E)$ are neglected, 
Eqs.~(3c) and~(3d) give
\begin{eqnarray}
\Omega'b_{1lr}(E)\approx\frac{(E+E_l-E_2-E_r)\Omega'^2b_1(E)
+\Omega'\Omega_0\Omega{b_2(E)}}{(E+E_l-E_1-E_r)(E+E_l-E_2-E_r)-\Omega_0^2}
\end{eqnarray}
%
Because the energy levels in each electron reservoir of the QPC are
dense, we can replace each sum over $l$ and $r$ in Eqs.~(3a)
and~(3b) by an integral. For instance,
\begin{math}\sum_{lr}{\rightarrow}{\int}{\int}{\rho}_l(E_l)
{\rho}_r(E_r)dE_ldE_r\end{math}, where $\rho_{l,r}$ is the density
of states in the left and right reservoirs, respectively. This
integral can be split into two parts: the principal and singular
value parts. Here the bandwidth of the QPC reservoirs are much larger than 
$V_d$ and the principal part is thus negligibly small~\cite{Mozyrsky02}. 
Actually, the principal part merely renormalizes the energy levels 
and the singular value part plays the dominant role. 
When Eq.~(A.1) is substituted into
$\sum_{lr}\Omega' b_{1lr}(E)$, we thus obtain two terms. The first
term is given by
\begin{eqnarray}
\sum_{lr}
\frac{(E+E_l-E_2-E_r)\Omega'^2b_1(E)}{(E+E_l-E_1-E_r)(E+E_l-E_2-E_r)-\Omega_0^2}
\nonumber\\
=b_1(E)\int_{-\infty}^{\mu_L}dE_l\int_{\mu_R}^{+\infty} dE_r
\frac{\rho_l(E_l)\rho_r(E_r)\Omega'^2(E+E_l-E_2-E_r)}
{(E+E_l-E_r-E_1')(E+E_l-E_r-E_2')}\nonumber\\
\approx{b_1(E)}\int_{-\infty}^{\mu_L}dE_l\int_{\mu_R}^{+\infty}dE_r
\big({-i\pi}\rho_l\rho_r{\Omega'}^2\big)\times
\nonumber\\
~~~\left\{\frac{E+E_l-E_r-E_2'}{(E+E_l-E_r-E_2')^2+\eta^2}\delta(E+E_l-E_r-E_1')
(E+E_l-E_2-E_r)\right.
\nonumber\\
~~~\left. +\frac{E+E_l-E_r-E_1'}{(E+E_l-E_r-E_1')^2+\eta^2}\delta(E+E_l-E_r-E_2')
(E+E_l-E_2-E_r)\right\}
\nonumber\\
=-i\frac{D'}{2V_d}b_1(E)
\left\{\frac{1}{\varepsilon'}(\frac{\varepsilon'}{2}+\frac{\varepsilon}{2})
(V_d+E-E_1')\theta(V_d+E-E_1')\right.
\nonumber\\
~~~\left. +\frac{1}{\varepsilon'}(\frac{\varepsilon'}{2}-\frac{\varepsilon}{2})
(V_d+E-E_2')\theta(V_d+E-E_2')\right\},
\end{eqnarray}
where~$V_d=\mu_L-\mu_R$, $D'=2\pi\rho_l\rho_r{\Omega'}^2 V_d$,
$\varepsilon=E_2-E_1$, $\varepsilon'=\sqrt{\varepsilon^2+4\Omega_0^2}$,
$E_{1,2}'=\frac{1}{2}(E_1+E_2)\mp\frac{1}{2}\varepsilon'$, and
$\theta(x)$ is the Heaviside step function.
In the high voltage limit with $eV_d\gg{\Omega}^2\rho$, the step
function in Eq.~(A.2) becomes one.
Thus, the L.H.S. of Eq.~(A.2) can be approximated by
$-i(D'/2V_d)(V_d+E-E_1)b_1(E)$.
When higher-order terms, i.e., the terms of the order $O(\Omega^6/V_d^2)$, 
are inculded, Eq.~(A.2) is approximated by 
$-i(D'/2V_d)\{V_d[1-\frac{1}{2}(D'/V_d)^2]+E-E_1\}b_1(E)$ 
in the high voltage limit. Obviously, the correction by $(D'/V_d)^2$ is very small 
for a large voltage $V_d$.

The second term is
\begin{eqnarray}
\fl \sum_{lr}
\frac{\Omega'\Omega\Omega_0b_2(E)}{(E+E_l-E_1-E_r)(E+E_l-E_2-E_r)-\Omega_0^2}
\nonumber\\
\fl = b_2(E)\int^{\mu_L}_{-\infty}dE_l\int_{\mu_R}^{+\infty}dE_r
\frac{\rho_l(E_l)\rho_r(E_r)\Omega'\Omega\Omega_0}{(E+E_l-E_1'-E_r)(E+E_l-E_2'-E_r)}
\nonumber\\
\fl \approx b_2(E)\int^{\mu_L}_{-\infty}dE_l
\int_{\mu_R}^{+\infty}dE_r (-i\pi\rho_l\rho_r\Omega'\Omega\Omega_0)
\left\{ \frac{1}{E+E_l-E_1'-E_r}\delta(E+E_l-E_2'-E_r)\right.
\nonumber\\
~~~~~~~~~~~~~~~~~~~~~~~~~~~~~~~~~\left.
+\frac{1}{E+E_l-E_2'-E_r}\delta(E+E_l-E_1'-E_r) \right\}
\nonumber\\
\fl =\frac{-i\Lambda}{2V_d}b_2(E) \frac{1}{E_2'-E_1'}
\left[(V_d+E-E_2')\theta(V_d+E-E_2')-(V_d+E-E_1')\theta(V_d+E-E_1')\right],
\nonumber\\
\end{eqnarray}
where $\Lambda=2\pi\rho_l\rho_r\Omega'\Omega_0\Omega V_d$.  In the
high voltage limit, the two step functions in Eq.~(A.3) become one
and the L.H.S. of Eq.~(A.3) is finally approximated by
$i({\Lambda}/{2V_d})b_2(E)$.  Thus, the sum of these two terms gives
Eq.~(\ref{eq4}) in the high voltage limit.  Substituting
Eq.~(\ref{eq4}) into Eq.~(3a), one obtains Eq.~(\ref{eq5a}).  
Also, one can derive Eqs.~(\ref{eq5b})-(\ref{eq5d}) using similar procedures. 
When higher-order terms, i.e., the terms of the order $O(\Omega^6\Omega_0/V_d^3)$,
are considered, Eq.~(A.3) is approximated by 
$i({\Lambda}/{2V_d})(1+\gamma)b_2(E)$ in the high voltage regime,
where the correction $\gamma=[(D+D')\sqrt{DD'}-DD']/4V_d^2$ is very small for 
a large voltage $V_d$.   



\section*{Acknowledgments} 

We would like to thank S. A. Gurvitz and Xinqi Li 
for valuable discussions. This work was supported by SRFDP, PCSIRT and 
the National Natural Science Foundation of China Grant
Nos.~10474013 and 10534060.

\end{document}